\documentclass[onecolumn]{svjour3} 
\usepackage{graphicx}
\usepackage{amsmath}
\usepackage{amssymb}
\usepackage{mathtools}
\usepackage{subfigure}
\usepackage{color}
\usepackage{subeqnarray}
\usepackage{mathrsfs}
\usepackage[colorlinks=true, pdfstartview=FitV, linkcolor=red,
  citecolor=blue, urlcolor=blue]{hyperref}
\usepackage{url}
\usepackage[utf8]{inputenc}
\usepackage{float}
\usepackage{siunitx}
\begin{document}
\title{Sliding and dry friction: Prandtl-Tomlinson athermal model revisited}
\author{María Luján Iglesias \and 
	Sebastián Gonçalves}
\institute{María Luján Iglesias \at
Instituto de Física, Universidade Federal do Rio Grande do Sul\\
	 Caixa Postal 15051\\
	 91501-970\\
	 Porto Alegre RS, Brazil.
	\email{lujaniglesias@gmail.com}
	\and
         Sebastián Gonçalves \at
\email{sgonc@if.ufrgs.br}
}

\date{\today}

\maketitle

\begin{abstract}
The microscopic origin of friction has been the goal of several
theoretical studies in the last decades. Depending on the investigated
systems or models, on the simulation techniques or conditions,
different and somewhat contradictory results have been found, even
when using the same model.  In this contribution we address this
apparent paradox in a well know case, the Prandtl-Tomlinson model at zero
temperature, studying the force-velocity relation for a wide range of
velocities not previously presented. Including much more data density
for the non trivial regions, we are able to shed light on this problem
and at the same time, provide new insight in the use of the
paradigmatic Tomlinson model for the secular problem of friction laws.
\end{abstract}


\section{Introduction}
What are the precise microscopic mechanisms that causes the appearance
of {\em friction forces} at the macroscopic level is one of the oldest
problems in physics, whose fundamental origin has been studied for
centuries and still remains
controversial~\cite{Persson2000,Krim2002,Makkonen2012}.  Our hominid
ancestors in Algeria, China, and Java (more than 400,000 years ago)
made use of friction when they chipped stone tools~\cite{persson2013},
for example.  Around 200,000 years ago, Neanderthals generated fire by
the rubbing of wood on wood or by the striking of flint stones.
Significant developments occurred some 5,000 years ago, as an Egyptian
tomb drawing suggests that wetting the sand with water to lower the
friction between a sled and the sand~\cite{Fall2014} was used for
moving large rocks.  The scientific formalization of such empirical
knowledge started with Da Vinci, followed by Amonton and Coulomb. They
established that the friction experienced by a body in contact with an
even surface is proportional to the load.  Second, the amount of
friction force does not depend on the apparent area of contact of the
sliding surfaces. And third, the friction force is independent of
velocity, once motion
starts~\cite{Baumberger1996,MuserWenning2001}. These three laws,
commonly verified in a macroscopic scale, are the result of the
collective behavior of many single asperity contacts, as was shown by
Bowden and Tabor(1954)~\cite{Bowden1951}.

With the introduction of the atomic force microscope
(AFM)~\cite{Binnig1986} and friction force microscope
(FFM)~\cite{Mate1987}, Bowden and Tabor's theory could be
experimentally verified, proving that friction laws for a single
asperity are different from macroscopic friction laws.  One of the
main results, confirmed by several
experiments~\cite{Mate1987,Fujisawa1995}, is that the friction force
on the nanometer scale exhibits a saw-tooth behavior, commonly known
as ``stick-slip'' motion. This observation can be theoretically
reproduced within classical mechanics using the Prandtl-Tomlinson
model~\cite{Tomlinson1929}.
 
Over many years, this model has been referred as the ``Tomlinson
model" even though the paper by Tomlinson did not contain it.  In
fact, it was Ludwing Prandtl who suggested in 1928 a simple model for
describing plastic deformation in crystals~\cite{Prandtl1928}. His
contributions were more associated with fluid
mechanics~\cite{Prandtl1904}, mechanics of plastic deformations,
friction, and fracture mechanics~\cite{Prandtl1920}.  In order to
correct this historical error, in 2003 Müser, Urbakh, and Robbins,
published a fundamental paper~\cite{Muser2003} in which the mentioned
model was termed ``Prandtl-Tomlinson Model"~\cite{Popov2014}.  Indeed,
the Prandtl-Tomlinson (PT) model has received some renewed attention,
as can be seen for example in modeling the aging effect on friction at
the atomistic scale~\cite{Mazo2017}.

On the other side, Makkonen~\cite{Makkonen2012}, using a thermodynamic 
approach, relates friction to the surface energy involved at the edges of 
nanoscale contacts between materials, as the result of new surface formation. 
This is a different approach as compared with the Prandtl-Tomlison model
presented here, which assumes that friction arises within the nanocontacts.

In the last years, theoretical
predictions for the atomic friction, based on the Prandtl-Tomlinson and
Frenkel-Kontorova~\cite{Kontorova:431596,Frenkel:431595,braun2004frenkel}
models, were proposed. The advantage of such models resides in being
simple and yet retaining enough complexity to show interesting
features.  Such models were able to explain some features of
atomic-scale friction, relating the energy dissipation with the
stick-slip motion, atomic vibration, and
resonance~\cite{Buldum1997,Goncalves2004,Goncalves2005,Fusco2005,Tiwari2008,Neide2010}.

In the original experiments of Mate et al.~\cite{Mate1987} the authors
state that the frictional force of a tungsten tip on graphite shows
little dependence on velocity for scanning velocities $v_{c}$ up to
\SI{400}{nm/s}.  A similar behavior has been reported in the work of
Zworner {\em et al.}~\cite{Zworner1998} for velocities up to several
\SI{}{\mu m/s}, where friction on different carbon structures has been
studied. They claim that a 1D Prandtl-Tomlinson model at $T = 0$ can reproduce
the velocity independent friction force for scanning velocities up to
\SI{10}{\mu m/s}, while giving rise to linear increase of friction for
higher velocities.  Other works claim a logarithmically increase in
the friction force with velocity, attributed to thermal
activation~\cite{Bouhacina1997,Bennewitz1999,Gnecco2000,Riedo2003,Li2011,Jinesh2008,Sang2001}.
Fusco and Fasolino~\cite{Fusco2005} have shown that an appreciable
velocity dependence of the friction force, for small scanning
velocities (from \SI{1}{nm/s} to \SI{1}{\mu m/s}), is inherent to the
Prandtl-Tomlinson model, having the form of a power-law $F_{fric} - F_{0}
\propto v_{c}^{2/3}$.  Considering the variety of seemingly
controversial results, we conducted the present study, producing a
wide range of numerical data for the friction force as a function of
the scanning velocity. With a higher density of points it is possible
to identify four clearly behavioral regions, which may go unnoticed
depending on how the data is presented.

In this way, we show the results in different scales of
representations to demonstrate how the conclusions appear to be
conflicting. At the same time, an overlooked region of data shows an
interesting behavior not previously reported.

\section{Methodology}
\begin{figure*}[!h]
\centering
\includegraphics[width=0.75\textwidth]{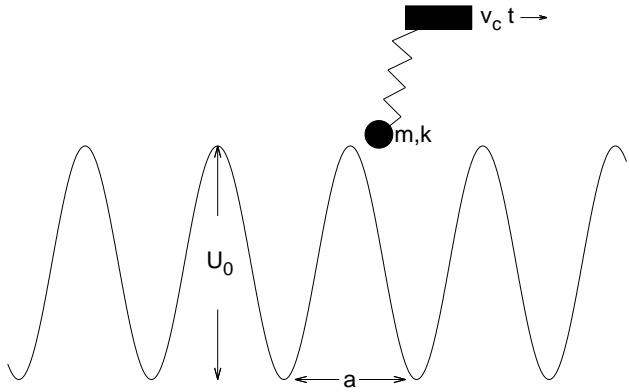}
\caption{Sketch of the 1D Prandtl-Tomlinson model for atomistic
  friction. The cantilever tip of mass $m$ and constant $K$ is moving
  at constant velocity $v_{c}$. The surface is represented as a
  potential with corrugation $U_{0}$ and period $a$ }\label{fig:toml}
\end{figure*}

We use the 1D Prandtl-Tomlinson model at $T = 0$ to simulate a tip of mass $m$
attached by a spring of constant $k$ to a support (cantilever) moving
at constant velocity $v_{c}$ along the $X$ direction, over a surface
represented by a periodic potential $V(x)$, where $x$
represents the position of the tip.  A graphical representation of the
model is shown in Fig.~\ref{fig:toml}.

The interaction potential has the form
\begin{equation}\label{potential}
V(x) = U_{0}\cos(\frac{2\pi x}{a}) \;,
\end{equation}
where $a$ is the lattice spacing. The elastic interaction between the
tip and the support is
\begin{equation}\label{elastic}
V_{el}(x) = \frac{1}{2}k(x - x_{c})^{2}\;,
\end{equation}
where $x_{c} = v_{c}t$ is the equilibrium position of the spring.
Thus, the equation of motion for this system, including the {\em
  ad-hoc} dissipation term, is
\begin{equation}\label{eq:Tomlinson}
m\frac{d^2 x}{dt^2} = -k(v_c t - x) +
U_{0}\frac{2\pi}{a}\sin(\frac{2\pi}{a}x) - m \gamma \frac{dx}{dt} \;,
\end{equation}

The therm proportional to the tip velocity $\frac{dx}{dt}$ 
is added to introduce energy dissipation in the model. Being $\gamma$
this proportionality constant.  Equation~\ref{eq:Tomlinson}
represent the same model used in some previous contribution to which
we want to make contact~\cite{Fusco2005,Zworner1998}.

The lateral force $F$ is calculated as $F= k (x_{c} - x)$, whereas
the frictional force $F_{fric}$ is identified as the lateral force
averaged over time $\langle F\rangle$~\cite{Tomanek1991}.  We solve
the nonlinear equation~\ref{eq:Tomlinson}, using the velocity Verlet
algorithm~\cite{VERLET1968} for a wide range of scanning velocity
$v_{c}$.

\section{Results}
In this section we present the results obtained by solving numerically
the equation of motion (Eq.~\ref{eq:Tomlinson}) for the
Prandtl-Tomlinson model.  The values of the constants for the model
are: $k$ = \SI{10}{N/m}, $m$ = \SI{e-10}{kg} (which gives a natural
frequency for the tip, $\sqrt{k/m} \simeq$ \SI{316}{kHz}), and $a$ =
\SI{0.3}{nm}, typical values of AFM
experiments~\cite{Bennewitz1999,Holscher1997,Zworner1998,Fusco2005}.
In general, the amplitude used for the corrugation $U_{0}$ goes from
0.2 to \SI{2}{eV}~\cite{Riedo2003}, and in the present case we use
$U_{0}$ = \SI{1}{eV}.  We chose $\gamma = 2\omega = 2\sqrt{k/m}$ in
order to have critical resonance of the system, and the time step used
for the numerical integration was $\Delta t \simeq$
\SI{1}{ns}~\footnote{The period of oscillation of the cantilever is
  $\SI{20}{\mu s}$ and the maximum simulated speed is $\SI{1}{mm/s}$,
  so that time step is more that 1000 times smaller that the period
  and at the maximum speed it moves only $1/300$ of the potential
  length}.  These particular set of parameters were chosen in order to
compare our results with those obtained by Zworner \textit{et
  al.}~\cite{Zworner1998} and Fusco and Fasolino~\cite{Fusco2005}.

\begin{figure*}[!h]
\includegraphics[width=0.5\columnwidth]{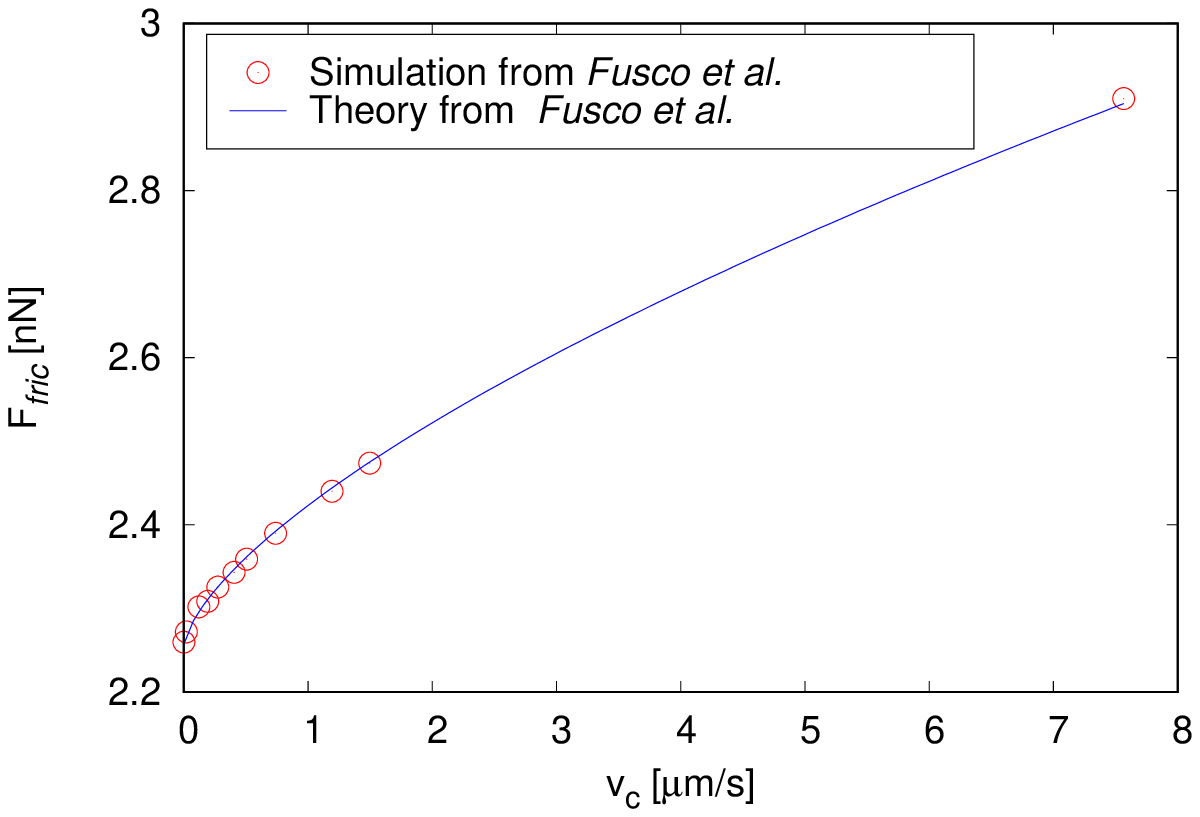}
\includegraphics[width=0.5\columnwidth]{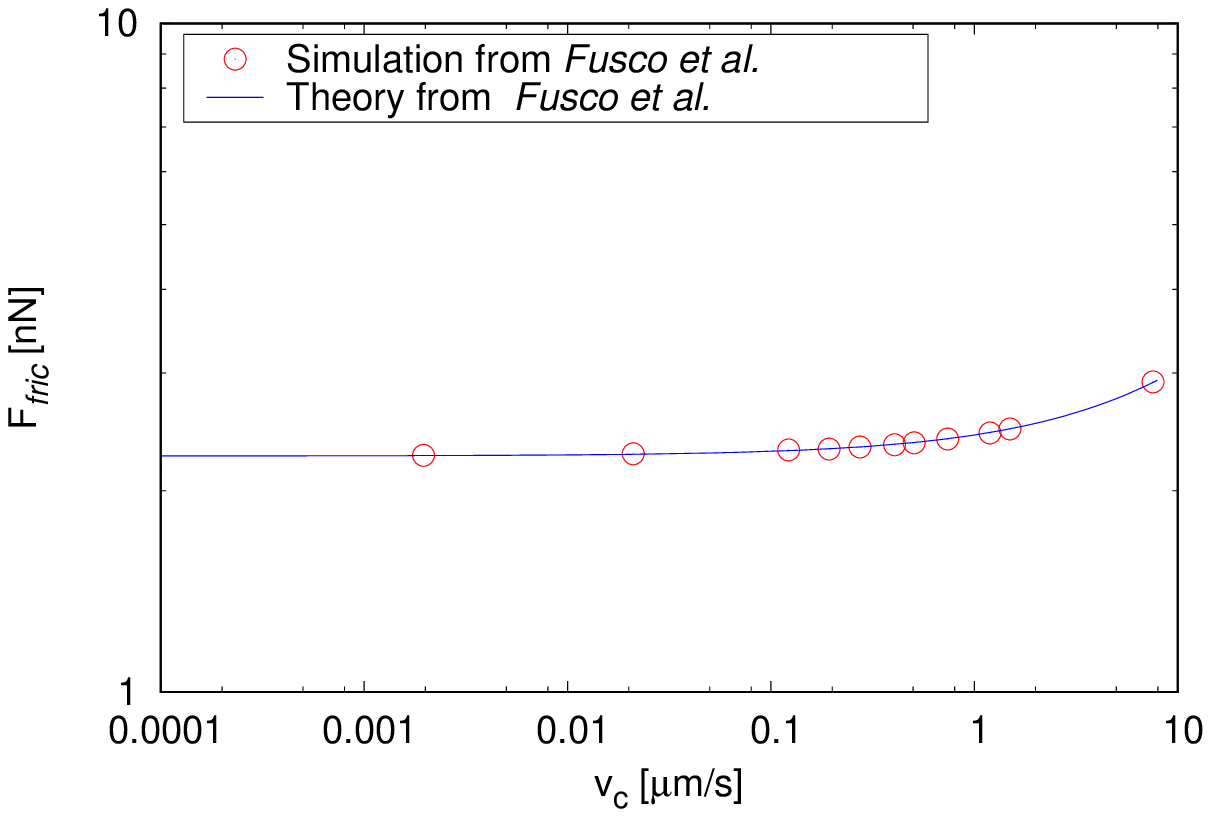}
\caption{Data extracted from Fusco and Fasolino~\cite{Fusco2005} showing the
  friction force ($F_{fric}$) as a function of the sliding
  velocity ($v_{c}$), plotted on a linear (top) and on a log-log
  scale(bottom), in the same way it was presented in the original article.}
\label{fig:fussco}
\end{figure*}

\begin{figure*}[!h]
\centering \includegraphics[width=0.5\columnwidth]{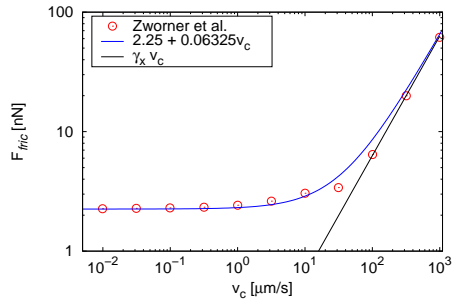}
\caption{Data extracted from Zworner \textit{et
    al.}~\cite{Zworner1998} showing the frictional force ($F_{fric}$)
  as a function of the sliding velocity ($v_{c}$), along with the
  analytic expressions proposed by them for the two limiting regimes.
}
\label{fig:zworner}
\end{figure*}

With the help of the software Engauge Digitizer~\cite{Engauge}
we recover the data points from the graphics of 
the mentioned references; in
this way we can reproduce the plots of their simulations in the most
similar way to the original articles.  Figure~\ref{fig:fussco} (left
and right) reproduces the results of Fusco and
Fasolino~\cite{Fusco2005} for linear and log-log scale, and
Fig.~\ref{fig:zworner} the corresponding ones of Zworner \textit{et
  al.}~\cite{Zworner1998}.  

\begin{figure*}[!h]
\includegraphics[width=0.5\columnwidth]{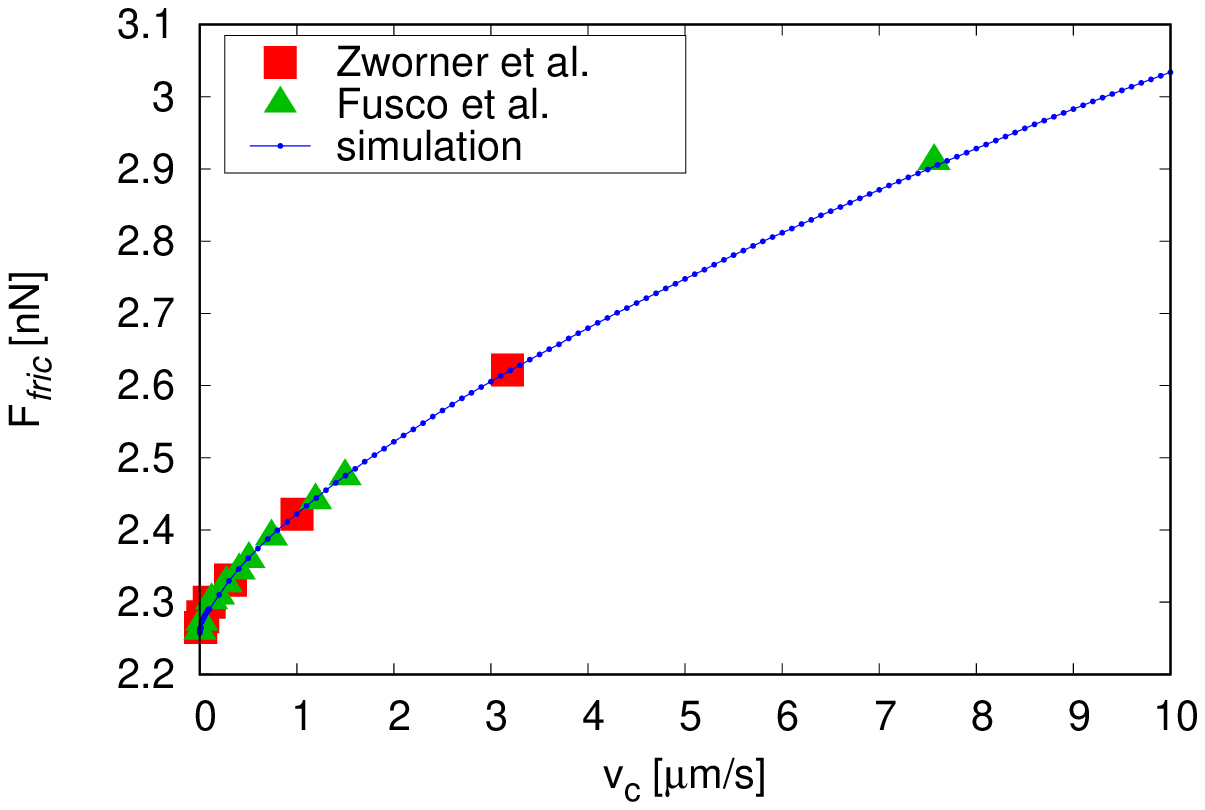}
\includegraphics[width=0.5\columnwidth]{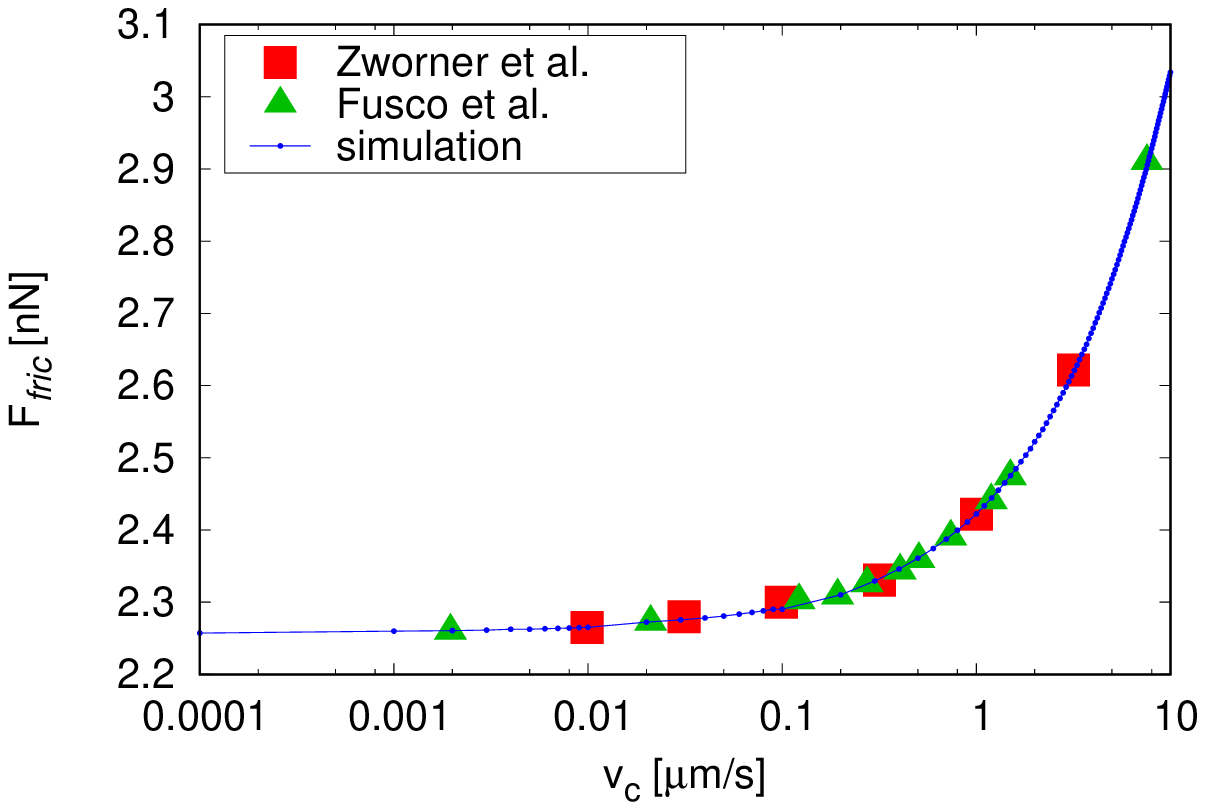}
\caption{$F_{fric}$ as a function of $v_{c}$ in semi-log scale;
  comparison between previous and present numerical results with the
  athermal Prandtl-Tomlinson model: Zworner \textit{et
    al.}~\cite{Zworner1998}, Fusco and Fasolino~\cite{Fusco2005}, and
  present contribution.}
\label{fig:comparacao}
\end{figure*}

Moreover, with the digitized data at hand, we can put the two sets in
the same graph along with the results from our own simulations (as
shown in Fig.~\ref{fig:comparacao}), so we can compare the three set
of data in the different regions.

Our numerical results are extended down to \SI{0.1}{nm/s}, in order to
show that, in principle, all data seems to be consistent. It
can be appreciated from the plot in linear scale
(Fig.~\ref{fig:comparacao} left), that this consistency is
hard to appreciate, particularly under $1\mu m/s$ where many order of
magnitude are condensed. For this reason, the same data has to be
presented in a more convenient scale, which is the case of a semi-log
scale (Fig.~\ref{fig:comparacao} right).

Zworner \textit{et al.}~\cite{Zworner1998} use a wide range of
velocities (from \SI{e-2}{\mu m/s} to \SI{e3}{\mu m/s}) so it is
interesting to analyze in detail the results on different regions.

\begin{figure*}[!h]
\centering
\includegraphics[width=0.5\textwidth]{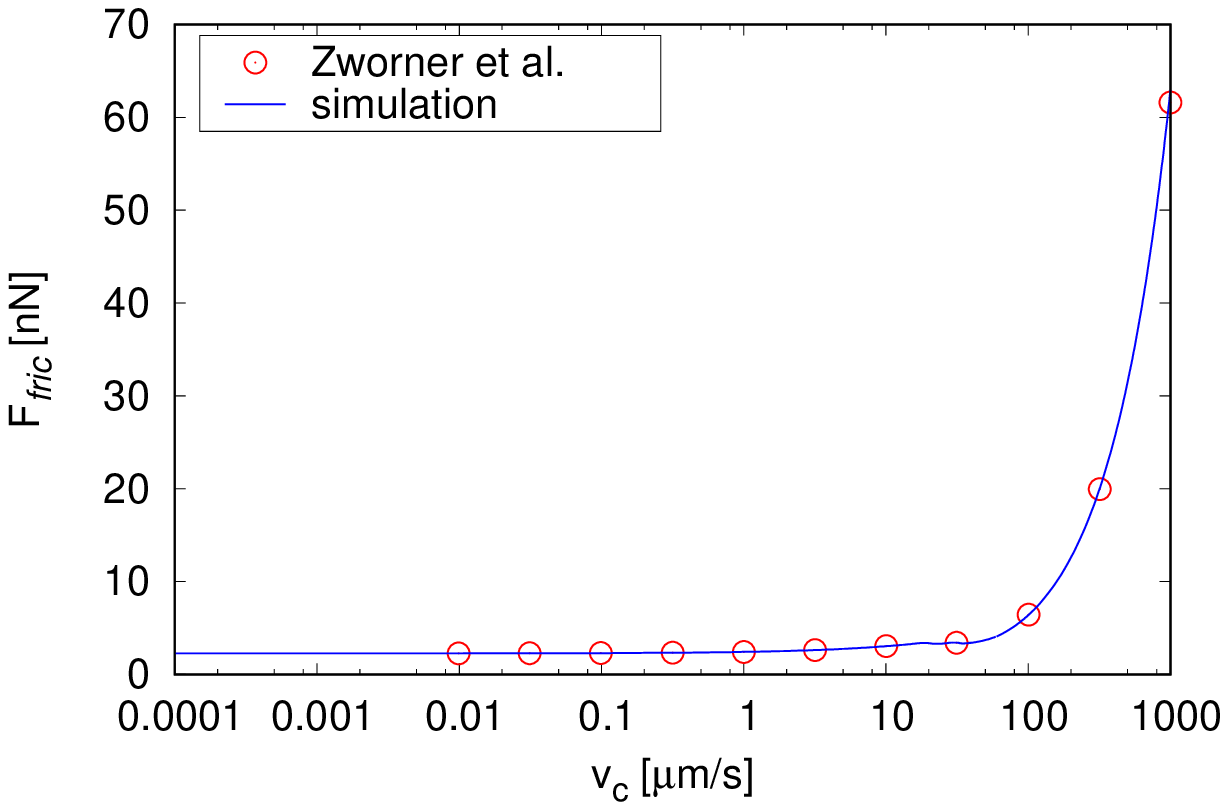}
\caption{$F_{fric}$ as a function of $v_{c}$ in linear and semi-log scale;
  comparison between Zworner \textit{et
    al.}~\cite{Zworner1998} simulations and present numerical results with the
  athermal Prandtl-Tomlinson model.}
\label{fig:comparison}
\end{figure*}

In their work, the authors concluded that the model exhibits two
limiting behaviors for the resulting frictional force
(Fig.~\ref{fig:zworner}): a velocity independent regime at low
velocities, below \SI{1}{\mu m/s} and a viscous linear regime for
velocities $\geq$ \SI{100}{\mu m/s}.  We can see in
Fig.~\ref{fig:comparison} that these two limiting cases ---and
particularly the combination of both--- can give an approximated
description of the Prandtl-Tomlinson model results in the wide region
of velocities displayed. However, one has to be aware that the log-log
scale used to present the data might hide possible departures from
such behavior, providing an oversimplification of the otherwise rich
features of the Prandtl-Tomlinson model that we will show in section
~\ref{subsec:slv}

\begin{figure*}[!h]
\includegraphics[width=0.5\columnwidth]{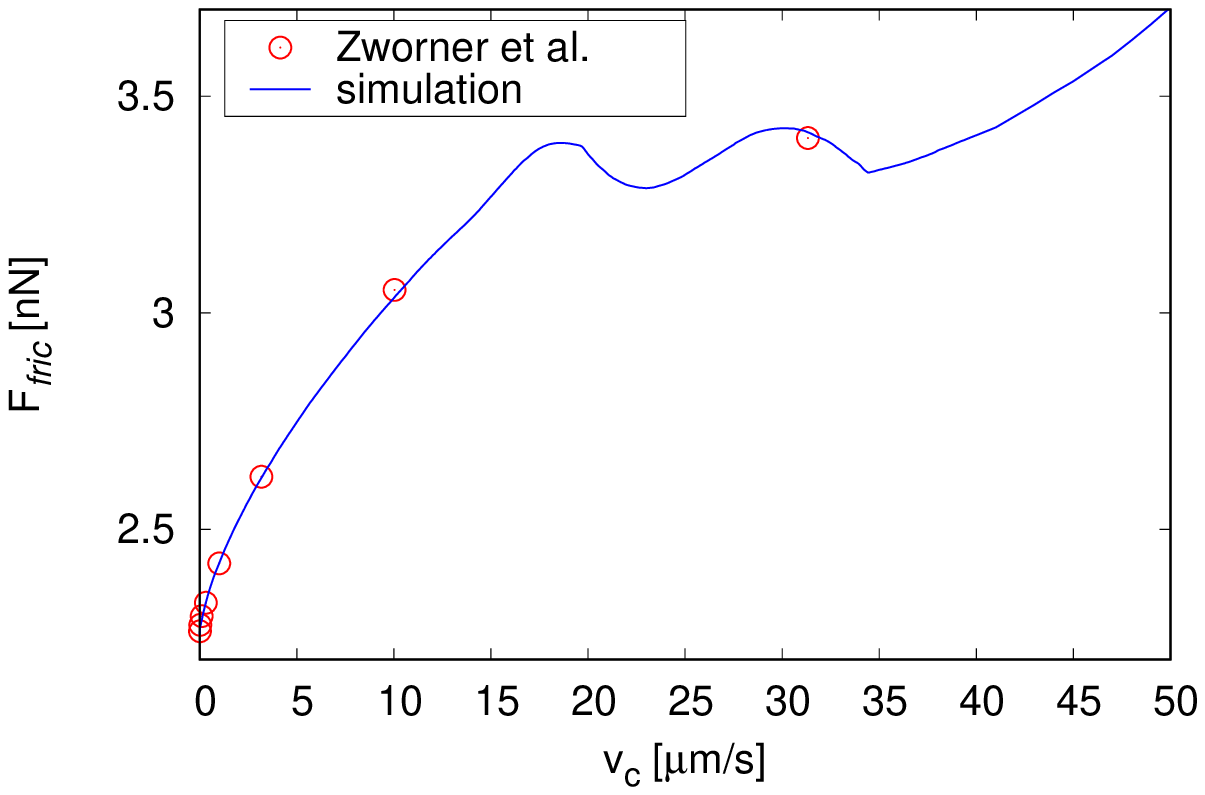}
\includegraphics[width=0.5\columnwidth]{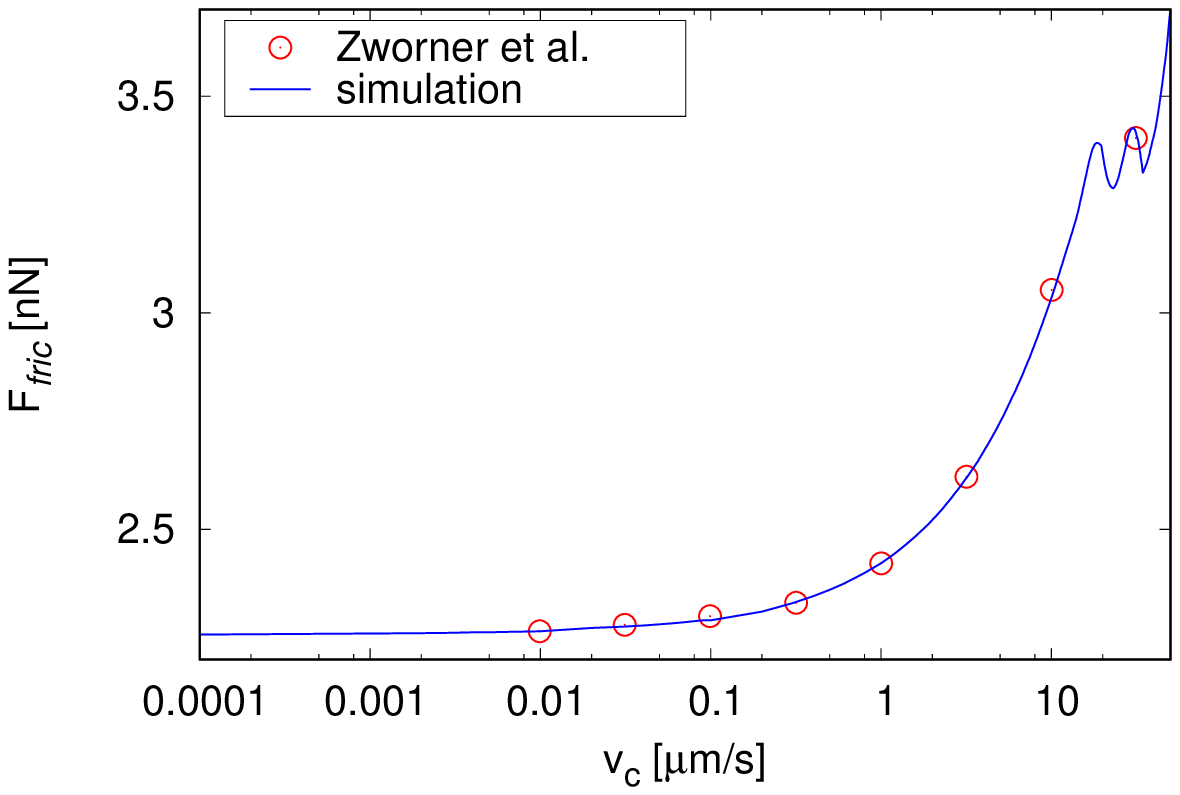}
\caption{Comparison between data from Zworner et
  al.~\cite{Zworner1998} and from present contribution in linear and semi-log scale.}
\label{fig:comparison2}
\end{figure*}

According with Zworner \textit{et al.} interpretation there is no
change in the frictional force when the velocity goes from
\SI{10}{nm/s} to \SI{1}{\mu m/s}. After this, $F_{fric}$ is
proportional to $\gamma v_{c}$ in a viscous damping regime.  Our
results were done in a wide interval of velocities, covering both
previous papers ranges and more, plus a small velocity increment.
Displaying the data in a linear scale (Fig.~\ref{fig:comparison2},
left), and with a higher density of points, we can see that the
dependence is far from being constant.  Moreover, on
Fig.~\ref{fig:comparison2} right, we show how the choice of the scale
can influence on the data interpretation, hiding relevant aspects of
the behavior of the system.

\subsection{Small Velocities representation}

\begin{figure*}[!h]
\centering
\includegraphics[width=0.5\textwidth]{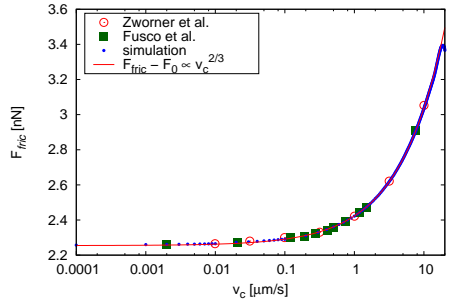}
\caption{Increase of the frictional force with velocity between
  \SI{e-3}{\mu m/s} to \SI{20}{\mu m/s}. The line is a power-law fit to the data
  of the form $F_{fric} - F_{0} \propto
  v_{c}^{2/3}$.~\cite{Fusco2005}}
\label{fig:ajuste1}
\end{figure*}

For small velocities (up to \SI{1}{\mu m/s}) we also make contact with the
results of Fusco and Fasolino~\cite{Fusco2005} by comparing the results of
our own simulations with the digitized data from them in 
Fig.~\ref{fig:ajuste1}.

In their article, the authors develop an approximation to the
dependence of the frictional force with the velocity, showing
that $F_{fric}$ follows a power law of the form:
\begin{equation}
F_{fric} = F_0 + c v_{c}^{2/3}
\end{equation}
where $c$ is a constant that depends on the parameter of the model
and on the space dimension.  This approximation is very accurate for
this range of velocities. Highlighting again that the friction is not
independent of the velocity. This proves that depending on the scale
chosen to represent the data, important information can be overlooked.
The same Figure also shows that this power-law dependency
continues to be valid for higher velocities of up to \SI{20}{\mu m/s}.
 
\subsection{Transitional and Large Velocities representation}
\label{subsec:slv}
\begin{figure*}[!h]
\centering \includegraphics[width=0.5\textwidth]{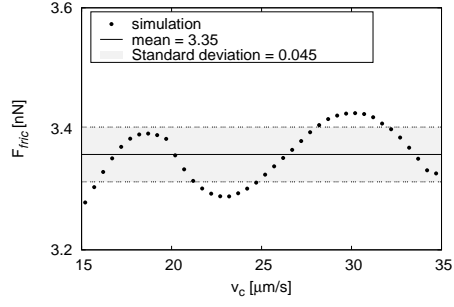}
\caption{Detail of the behavior of the friction force with velocity
  between \SI{15} to \SI{35}{\mu m/s}. We can say that friction is
  stationary in this range of velocities.}
\label{fig:ajuste3}
\end{figure*}

An important fact that stands out is the distinctive behavior in the
region between \SI{10}{\mu m/s} and \SI{35}{\mu m/s}
(Fig.~\ref{fig:ajuste3}), where $F_{fric}$ oscillates with $v_c$
around a fixed value of force. Indeed, for the velocities $v_{c}$ in
this region,the tip sees a force from the surface that varies with the
a frequency close to its own natural frequency~\cite{Goncalves2005,Apostoli2017},
{\em i.e.}:
\begin{equation}
  v_c^{res} = \frac{a}{2\pi}\sqrt{\frac{k}{m}} \simeq \SI{15}{\mu m/s}
\end{equation}

\begin{figure*}[!h]
	\centering
        \includegraphics[width=0.5\textwidth]{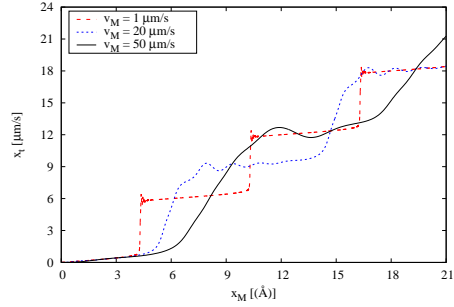}
	\caption{Tip position as a function of cantilever position for
          different sliding velocities, before, at, and after the
          region where friction oscillates.}
	\label{fig:tip-pos}
\end{figure*}

\begin{figure*}[!h]
	\centering
        \includegraphics[width=0.5\textwidth]{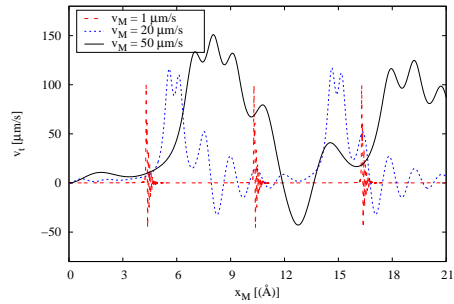}
	\caption{Tip velocity as a function of cantilever position for
          different sliding velocities, before, at, and after the
          region where friction oscillates.}
\label{fig:tip-vel}
\end{figure*}

In order to better understands what could be the origin of the
reported behavior we present below the tip position and velocity as a
function of the cantilever position (Fig.~\ref{fig:tip-pos}). We
present that data for different values of the sliding cantilever
velocity. For velocities well below the region where friction
oscillates, we have the typical stick-slip behavior where energy
dissipates mainly after the slip movement. For large velocities the
viscous regime is recovered.  However in the region of interest we
observe that the tip slips over a distance of two cell parameters,
from one potential barrier to the second other one. This happens only
for a damping coefficient near critical. Then, instead of being a 
resonance phenomenon, it is a critical damping effect where the 
dissipation remains at its minimum compatible with the sliding 
speed (Fig.~\ref{fig:tip-vel}).
The presence of double jumps for that range of velocities, makes us to wonder if multiple (more than two) slide jumps would be possible. That can happen, depending on the key parameters of the model, {\em i.e}, 
$U_0$, $k$, and $a$~\cite{Gnecco2012}, and it deserves a future 
and thorough analysis.

Such behavior comes to light in reason of the remarkable increase in
the density of simulated points and it has not been reported before.  In
this segment the value of $F_{fric}$ oscillates around an average
value of \SI{3.35}{nN}. This behavior goes totally unnoticed in a
logarithmic representation of the data. Some recent
experiments~\cite{Chen2006,Diao2018} have been carried out evaluating
the friction dependence on speed near this region, 
but without covering the entire range and without sufficient detail and precision.

\begin{figure*}[!h]
\centering
\includegraphics[width=0.5\textwidth]{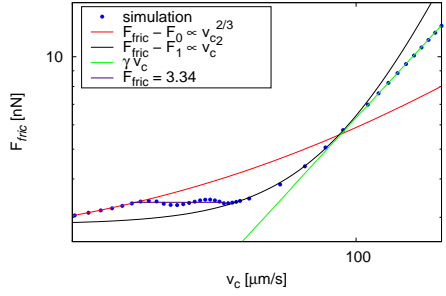}
\caption{The friction force as a function of velocity in the transition region
  between the low velocities $v_{c}^{2/3}$ regime to the high velocities,
  viscous linear regime.  For that intermediate regime of
  velocities the force goes through two other transitional regimes of
  almost constant force to quadratic velocity regime before entering
  the linear regime.  Increase of the frictional force with velocity
  between \SI{35} to \SI{100}{\mu m/s}. The line is a power-law fit to
  the data of the form $F_{fric} - F \propto v_{c}^{2}$. For high
  velocities the frictional force is proportional to the velocity in
  the regime of viscous damping}
\label{fig:ajuste4}
\end{figure*}

After the transition region, we observe that friction growths again.
We fit the data with a quadratic power-law:
\begin{equation}
F_{fric} - F \propto v_{c}^{2}
\end{equation}
where $F$ is the value of the friction at the lower velocity of the
adjustment (Fig.~\ref{fig:ajuste4}). The quadratic growth fits very
well with the simulation data up to \SI{0.1}{mm/s} behaving like a
drag force until finally, for large sliding velocities, the mechanism
of energy dissipation through the ``stick-slip'' effect breaks down,
and $F_{fric}$ is proportional to $\gamma v_{c}$, where critical
damping is assumed, giving $\gamma = 2 \sqrt{k m}$.

\section{Conclusion}
We have presented a thorough numerical study on the velocity
dependence of friction that emerge from the classical
Prandtl-Tomlinson model in the athermal case.  Despite the fact that
similar works have already been carried out, our contribution was made
with a higher density of points, bringing to light behaviors not
previously reported.  By comparing our results with previous ones with
the same model, we were able to conciliate apparent conflicting results
while providing new insight and interpretation of them.  Besides, we
present results in regions not previously explored.  We confirm Fusco
and Fasolino results for small velocities but extended up to
\SI{15}{\mu m/s}, where friction force has a dependence of
$v_{c}^{2/3}$.  A transition region located between
\SIrange{15}{35}{\mu m/s} where there is a constant frictional force
on average and then an increase proportional to $v_{c}^{2}$ up to
\SI{100}{\mu m/s}. After this, the force is proportional to $\gamma
v_{c}$ in a viscous damping regime.  Our numerical study shows that
depending on how the results are presented, mainly when changing from
linear scale to logarithmic, part of this rich an interesting behavior
can go unnoticed.

\begin{acknowledgements}
This work was supported by the Centro Latinoamericano de Física
(CLAF), the Conselho Nacional de Desenvolvimento Científico e
Tecnológico (CNPq, Brazil) and in part by the Coordenação de
Aperfeiçõamento de Pessoal de Nível Superior - Brasil(CAPES) - Finance
Code 001.
\end{acknowledgements}

\bibliographystyle{spbasic}

\end{document}